# Approaching Three-Dimensional Quantum Hall effect in Bulk HfTe$_5$

Pang Wang[1,2,#], Yafei Ren[1, #], Fangdong Tang[2,3], Peipei Wang[2], Tao Hou[1], Hualing Zeng[1,†], Liyuan Zhang[2,†], Zhenhua Qiao[1,†]

1 International Center for Quantum Design of Functional Materials, Hefei National Laboratory for Physical Sciences at the Microscale, Synergetic Innovation Centre of Quantum Information and Quantum Physics, CAS Key Laboratory of Strongly Coupled Quantum Matter Physics, and Department of Physics, University of Science and Technology of China, Hefei, China.

2 Department of Physics, Southern University of Science and Technology, and Shenzhen Institute for Quantum Science and Engineering, Shenzhen, 518055, China

3 Solid State Nanophysics, Max Plank Institute for Solid State Research, Stuttgart, Germany.

[#] These authors contribute equally to this work.
[†] hlzeng@ustc.edu.cn, zhangly@sustech.edu.cn, and qiao@ustc.edu.cn

The discovery of quantum Hall effect in two-dimensional (2D) electronic systems inspired the topological classifications of electronic systems[1,2]. By stacking 2D quantum Hall effects with interlayer coupling much weaker than the Landau level spacing, quasi-2D quantum Hall effects have been experimentally observed[3~7], due to the similar physical origin of the 2D counterpart. Recently, in a real 3D electronic gas system where the interlayer coupling is much stronger than the Landau level spacing, 3D quantum Hall effect has been observed in ZrTe$_5$[8]. In this Letter, we report the electronic transport features of its sister bulk material, i.e., HfTe$_5$, under external magnetic field. We observe a series of plateaus in Hall resistance $\rho_{xy}$ as magnetic field increases until it reaches the quantum limit at 1~2 Tesla. At the plateau regions, the longitudinal resistance $\rho_{xx}$ exhibits local

**minima. Although $\rho_{xx}$ is still nonzero, its value becomes much smaller than $\rho_{xy}$ at the last few plateaus. By mapping the Fermi surface via measuring the Shubonikov-de Haas oscillation, we find that the strength of Hall plateau is proportional to the Fermi wavelength, suggesting that its formation may be attributed to the gap opening from the interaction driven Fermi surface instability. By comparing the bulk band structures of ZrTe$_5$ and HfTe$_5$, we find that there exists an extra pocket near the Fermi level of HfTe$_5$, which may lead to the finite but nonzero longitudinal conductance.**

*Introduction---.* Two benchmarks of the quantum Hall effect (QHE) are the quantized Hall resistance plateaus and vanishing longitudinal resistance[9]. The quantization originates from the presence of energy gap, which appears once the Landau levels are formed under strong magnetic field[10]. By stacking layers of two-dimensional (2D) QHE into a superlattice, the signatures of QHE can also be observed in bulk systems when the interlayer coupling becomes smaller than the Landau level specings[11]. Such bulk QHE is regarded as quasi-2D as its nature is still closely related to the 2D case[12]. Recently, the evidence of genuine 3D QHE has been experimentally reported by electronic transport measurements in bulk ZrTe$_5$ materials under magnetic field[8] after being proposed theoretically over thirty years ago by Halperin[13]. Contrast to the quasi-2D case, the single-particle electronic bands under the magnetic field are gapless since the interlayer coupling is much larger than the Landau level spacing. An energy gap near the Fermi energy is evidenced by transport measurements originating from the formation of charge density wave driven by Coulomb interaction[8,14].

Comparing to 2D QHE that is universal and can be realized in a large number of 2D electronic systems, the material systems hosting 3D QHE are extremely limited. In this Letter, we report our experimental findings suggesting the approaching of 3D QHE in bulk HfTe$_5$ systems. We first demonstrate the 3D nature of the Fermi surface via Shubnikov-de Haas oscillations of longitudinal resistances by changing the orientation. We then measure the Hall and longitudinal resistances by increasing the magnetic field to the quantum limit where Landau levels are formed. We find quasi-plateaus in Hall resistance and corresponding dips of longitudinal resistance, which together indicate the possible presence of 3D QHE. We show that the magnitude of the last Hall resistivity plateau $\rho_{xy}$ is proportional to the Fermi wavelength $\lambda_{F,z}$ along z-direction, i.e., $\rho_{xy} = \frac{h}{e^2}\frac{\lambda_{F,z}}{2}$, where $h$ is the Plank's constant and $e$ is an elementary charge, following the same rules as those findings in ZrTe$_5$[8]. These qualitatively similar results in HfTe$_5$ as those in ZrTe$_5$ suggest the same physical origin underlying both cases, i.e., a 3D QHE induced by Fermi surface instability driven by Coulomb interaction. Nevertheless, one can see that though the longitudinal resistance is much smaller than the corresponding Hall resistance, it does not completely decrease to be vanishing at the last plateau region. We attribute the possible physical origin to the emergence of multi-band around the Fermi surface of HfTe$_5$.

HfTe$_5$ is an orthorhombic layered structure composed of 1D atomic chains along *a*-direction and the layer stacking along *b*-direction as illustrated in Fig. 1a[15], where *a*, *b*, and *c* directions are set to be along *x*, *z*, and *y* axes. This material is topologically same as ZrTe$_5$, which is either considered as strong topological insulator or weak topological insulator[16~21], depending on the detailed materials. We measure the longitudinal and Hall resistances of bulk HfTe$_5$ systems with thicknesses of tens of micrometers at different temperatures ranging from 1.5 to 200 K by using the six-electrode Hall-bar geometry as schematically plotted in the Inset of Fig. 1b.

The longitudinal resistivity $\rho_{xx}$ of HfTe$_5$ exhibits an anomalous dependence on temperature as displayed in Fig. 1b, where $\rho_{xx}$ first increases and then decreases as temperature increases with a peak that appears at $T_p$ of 80.6 Kelvin. By measuring the Hall effect at different temperatures as displayed in Fig. 1c, we find that the carrier type is electron-like when $T<T_p$, whereas it becomes hole-like when $T>T_p$. This phenomenon is consistent with the Angle resolved photoemission spectroscopy (ARPES) results, i.e., the Fermi energy transitions from valence band to conduction band when temperature decreases[22].

*3D Fermi Surface ---.* After characterizing the samples in our experiments, we move to measure the Hall effect in our samples. We first try to detect the 3D nature of the electronic structure, i.e., a closed Fermi surface. By measuring the dependence of longitudinal resistivity $\rho_{xx}$ as a function of magnetic field, we find the presence of Shubnikov-de Haas oscillation for the magnetic field along with any orientation (See Section I of Supplementary Materials for the detailed results). In particular, as displayed in Fig. 1d, we plot the longitudinal resistance as a function of the parallel magnetic field with the electric current. We show that, at temperatures up to 7 Kelvin, the clear Shubnikov-de Haas oscillation still appears. This suggests that the electrons' motion in x-z or y-z plane is coherent and the Fermi surface is closed[8]. And we also find an obvious negative magnetoresistance which could be attributed to the chiral anomaly observed in many other materials[23~29].

***3D QHE at 1.5 Kelvin ---.*** We then focus on the electronic transport properties at a low temperature of 1.5 Kelvin. From both the longitudinal and Hall resistances at low magnetic field, we extract the carrier densities and mobilities of our samples, which are respectively $10^{16}$~$10^{17}$ cm$^{-3}$ and 100,000~146,000 cm$^2$V$^{-1}$S$^{-1}$ (See Section II of Supplementary Materials for the detailed fitting results). One can see that our samples exhibit very low carrier density and extremely high mobility. By applying the magnetic field along z direction, we observe the plateaus of Hall resistance $\rho_{xy}$ and oscillations of longitudinal resistance $\rho_{xx}$ when the applied magnetic field increases, as displayed in Fig. 2a. In the low-field region in Fig. 2b, we observe a series of oscillations of $\rho_{xx}$. The Landau filling factor *N* can be extracted from the Landau fan diagram as displayed in Fig. 2c, where the peak positions of $\rho_{xx}$ are marked by integers, and the valleys are marked by half integers. We find that the quasi-plateaus of $\rho_{xy}$ correspond to the filling factors from $N = 2$ to $N = 1$ as labeled by numbers. Moreover, we find that at about B = 1.8 Tesla the system reaches the extreme quantum limit, i.e., only the lowest Landau band being occupied. Qualitatively, our results are quite similar to that of 2D QHE, i.e., in the dip regimes of $\rho_{xx}$, $\rho_{xy}$ exhibits plateaus. However, quantitatively, there are some differences.

The magnitude of the last Hall resistance plateau is about 0.7 Ω that is much smaller than 25.8 kΩ in 2D QHE, strongly suggesting that the Hall resistance plateau in our studies originates from the 3D bulk system but not from the 2D system. To better reflect the 3D nature of the Hall effect, we express the Hall resistivity to be $\rho_{xy} = \frac{h}{e^2}\lambda_D$ following Halperin's work[13], which indicates that the sample with a thickness of $\lambda_D$ contributes to a quantized Hall conductance of $\frac{e^2}{h}$ at the $N$=1 plateau. In this manner, we can assume the bulk system under magnetic field as a superstructure with many supercells and the thickness $\lambda_D$ along the field direction. By comparing $\lambda_D$ with the Fermi wavelength $\lambda_{F,Z}$ along the field direction, i.e., $z$ direction, we find that the ratio of $\frac{\lambda_D}{\lambda_{F,Z}}$ is always about 1/2 for all three samples as displayed in Fig. 2d and Table I (See Section I of Supplemental Materials for details). Such a relation following the same rule as demonstrated in Ref. [8] suggests the presence of an energy gap at the Fermi energy due to the possible formation of charge density wave state[14,30,31]. Nevertheless, the nonzero longitudinal resistance preserves indicating that the gap might be smeared by disorder or there are other conducting electrons near the Fermi level contributing to finite conductance as discussed below.

To explore the possible underlying physical origin of nonzero $\rho_{xx}$, we first measure the dependence of the lowest $\rho_{xx}(B)$ at lower temperature as shown in Fig. S4 in Supplemental Materials. We find that the minimal magnitude of longitudinal resistance $\rho_{xx}$ shows weak dependence on temperature and saturates as temperature decreases, which indicates that the lowest value of $\rho_{xx}$ is not limited by the temperature. By further comparing the relative magnitude of $\rho_{min}/\rho_{xx}(B=0)$ for different samples with different mobilities, we notice that as mobility increases, $\rho_{min}/\rho_{xx}(B=0)$ shows the trend to decrease as shown in Fig. S5 of Supplemental Materials.

Comparing to the extremely large magnitudes of that in ZrTe$_5$ samples, i.e., 5E$^5$ cm$^2$V$^{-1}$S$^{-1}$, our best HfTe$_5$ sample exhibits mobility of about 2E$^5$ cm$^2$V$^{-1}$S$^{-1}$. Thus, one possible reason for the presence of finite $\rho_{min}$ is that the mobilities of our HfTe$_5$ samples are yet not high enough. Alternatively, another possible reason lies in the difference between band structures of HfTe$_5$ and ZrTe$_5$. As displayed in Fig. 3, we find one more valley appears at the Fermi level near the M point in HfTe$_5$, whereas the corresponding band around M is much higher than the Fermi energy in ZrTe$_5$. Both experimental and theoretical results indicate that the lattice constants of both materials would decrease as temperature decreases[32]. Our calculations by WIEN2K suggest that the valley near M point will become lower than that near Γ point in HfTe$_5$ when the lattice constant decreases, whereas in ZrTe$_5$, the valley near Γ point is always the lowest one as displayed in Fig. 3 (See more details in Section VI of Supplemental Materials). Such kind of difference between ZrTe$_5$ and HfTe$_5$ leads to different carrier types that contribute to the longitudinal conductance. For the ZrTe$_5$ case, only the topological bands near Γ point contribute to the electronic transport, while for the HfTe$_5$ case in this Letter, the electrons at M valley may also involve in the electronic transport, resulting in the nonzero longitudinal resistance when the band around Γ enters the extreme quantum limit region.

*Metal-Insulator Transition under High Magnetic Field---.* When we further increase the magnetic field to the extreme quantum limit, the longitudinal and Hall resistances stop oscillating as displayed in Fig. 2a. The longitudinal resistance $\rho_{xx}$ exhibits an unsaturated magnetoresistance, which can reach 10000% of the zero field resistance at the magnetic field of 13 Tesla. Simultaneously, the Hall resistance $\rho_{xy}$ keeps increasing to saturation in a non-linear manner. By varying temperatures, we find that the oscillation of Hall resistance becomes smoother, whereas the turning point at the last resistance plateau is quite robust up to the temperature of about 12 Kelvin as displayed in Fig. 4a. Moreover, we also find a metal-insulator transition at the extreme quantum limit as displayed in Fig. 4b, which is characterized by the longitudinal resistance at a critical transition point of $B_C \approx 7.07$ Tesla. From Fig. 4b, we can observe that, at $B<B_C$, $\rho_{xx}$ increases as temperature increases, which behaves like a metal; whereas at $B>B_C$, $\rho_{xx}$ decreases as temperature increases, which behaves like an insulator. Near the phase transition point, $\rho_{xx}$ is weakly dependent on the temperature as displayed in Fig. 4c. To further characterize the phase transition, we perform scaling analysis of $\rho_{xx}$ against the scaling variable $(B - B_C)T^{-1/\zeta}$ where $\zeta$ is the fitting parameter as displayed in Fig. 4d. We find that $\zeta \cong 15$, which does not agree with the metal-insulator transition in 2D QHE either with or without spin degeneracy of Landau levels[34]. Moreover, by changing the temperature range, we find different values of $\zeta$ [See more details in Sec. IX of the Supplementary Materials]. Thus, the phase transition cannot fit to the universality classes in 2D QHE and the nature of the insulating phase deserves further exploration in the future.

***Summary---.*** We report that bulk HfTe$_5$ (the only cousin material of ZrTe$_5$) in the presence of magnetic field can approach the 3D QHE with quantized Hall conductance and finite longitudinal resistance at the extreme quantum limit. By applying a magnetic field perpendicular to the atomic layers of bulk HfTe$_5$ and increasing it to the quantum limit, we observe Hall resistance plateaus and the corresponding longitudinal resistances show corresponding dips. After comparing the thickness of each supercell that contributes to a quantum Hall conductivity $\lambda_D$ and the Fermi wavelength $\lambda_{F,Z}$, we find that $\frac{\lambda_D}{\lambda_{F,Z}}$ is about 1/2 for different samples, suggesting that the presence of an energy gap. Nevertheless, the nonzero longitudinal resistance preserves that might be attributed to the presence of other conducting electrons near the Fermi level through our numerical calculation. Future experiments are still desired by using, for example, ARPES to demonstrate the electronic structure near the Fermi level[22,33].

## Acknowledgements

The authors thank Prof. Hongming Weng for valuable discussions. This work was supported by National Key R & D Program (2017YFB0405703, 2017YFA0205004 and 2018YFA0306600), and the NNSFC (11974327, 11474265, 11674295 and 11674024), the Fundamental Research Funds for the Central Universities (WK2030020032, WK2340000082), and Anhui Initiative in Quantum Information Technologies. This work was partially carried out at the USTC Center for Micro and Nanoscale Research and Fabrication.

**Table I:**

| sample | S#1 | S#2 | S#3 |
|---|---|---|---|
| $\lambda_{F,Z}(nm)$ | 6.5±0.2 | 13.1±0.6 | 3.3±0.2 |
| $\lambda_Q(nm)$ | 3.4±0.5 | 5.9±0.1 | 1.7±0.2 |
| $\lambda_Q/\lambda_{F,Z}$ | 0.52±0.1 | 0.45±0.03 | 0.52±0.1 |
| $\lambda_{F,Z}/0.725nm$ | 9±0.2 | 18±0.9 | 4.5±0.3 |

**Table I** | Relation between Fermi wavelength and period $\lambda_Q$ for 3 different samples. The detail method to get $\lambda_{F,Z}$ is discussed in supplementary materials and $\lambda_Q$ is calculated by the last plateau according to the equation $\rho_{xy,n} = \frac{h}{e^2}\lambda_Q$.

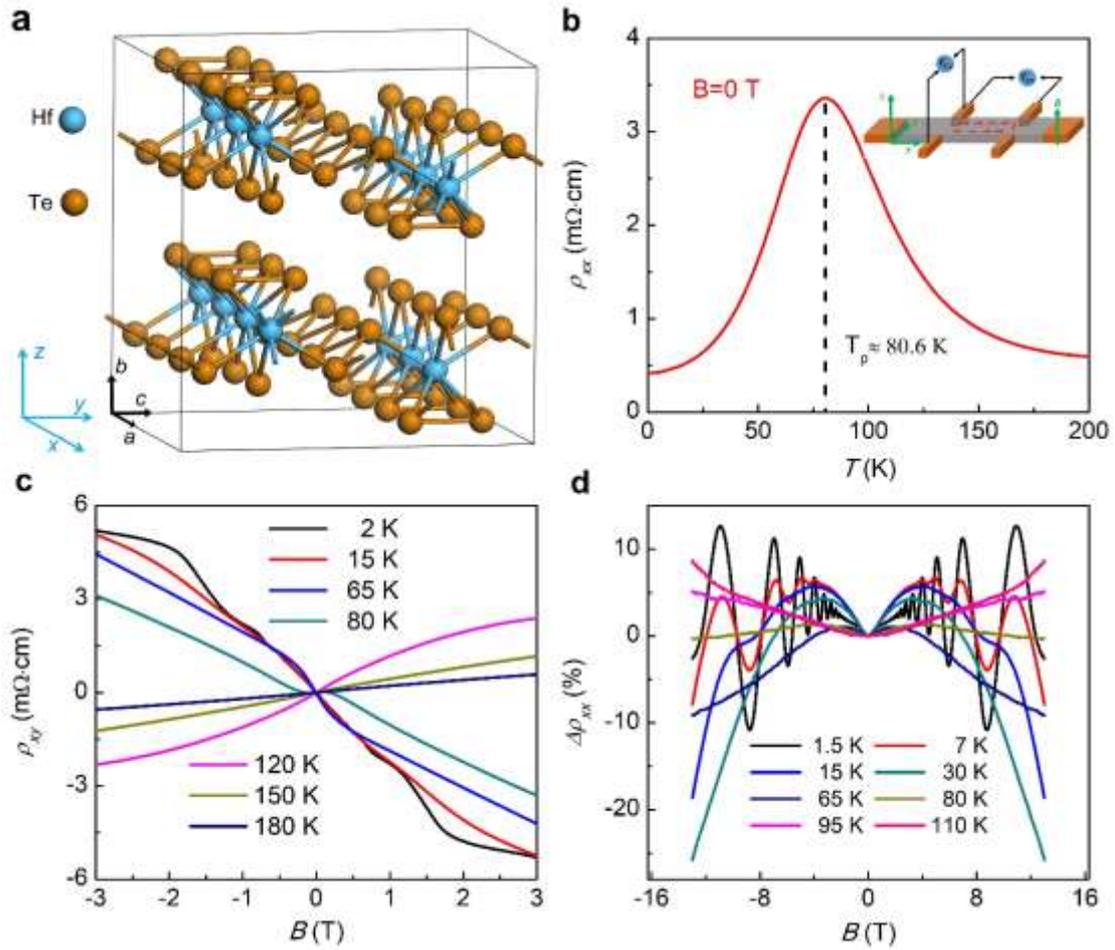

**Figure 1 | Basic structure and three-dimensional nature of HfTe5. a,** Crystal structure of HfTe5. **b,** Temperature-dependence of resistivity of bulk HfTe5. There is a peak at about 80.6 K labeled by dash line. The inset illustrates the setup of our transport measurement. **c**, Magnetic field-dependence of hall resistivity at different tempratures. When it higher than 80 K, the sign just changes which means the main carrier transits from electron to hole. **d**, The parallel magnetic field-dependence of resistivity ($\Delta\rho_{xx} = \frac{\rho_{xx}(B)-\rho_{xx}(0)}{\rho_{xx}(0)}$) at different temperatures.

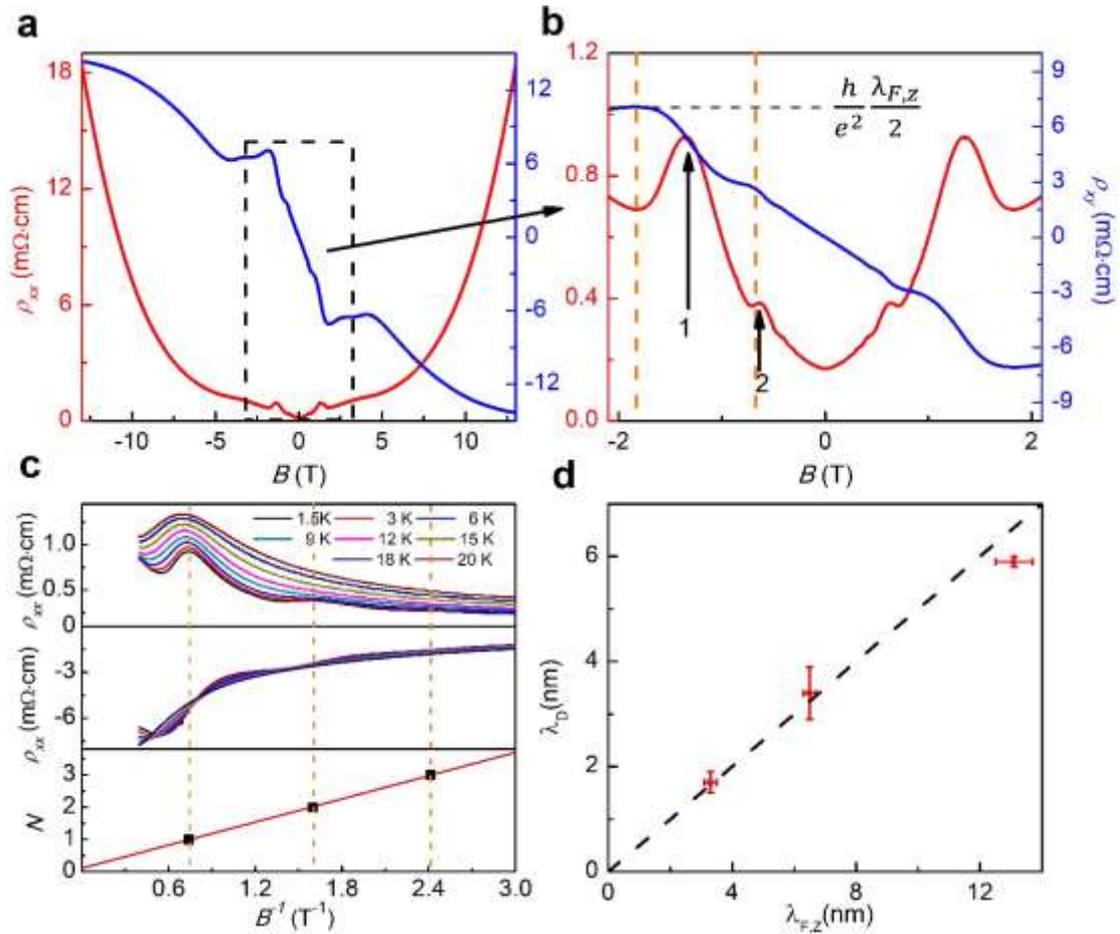

**Figure 2 | Signatures and analysis of three-dimensional quantum Hall effect.**
**a,** Out-of-plane perpendicular magnetic field-dependence of londitudinal (red) and Hall (blue) resistivity from 13 T to -13 T at 1.5 K. **b,** The zoomin of the dash frame in a. We can find that every Hall plateau just corresponds to the dip of the $\rho_{xx}$ marked by yellow dash lines. The black dash line marks the last Hall plateau whose resisticity equals to $\frac{h}{e^2}\frac{\lambda_{F,z}}{2}$. The arrows and numbers refer to the Landau-factor extracted by c. **c,** The $\rho_{xx}$ and $\rho_{xy}$ versus $1/B$ at different temperatures and the Landau fan diagram. The yellow dash lines mark the position of peak of $\rho_{xx}$. **d,** Period $\lambda_D$ versus z-direction Fermi wavelength $\lambda_{F,Z}$, the ratio is almost about 0.5 which is fitted by dash line.

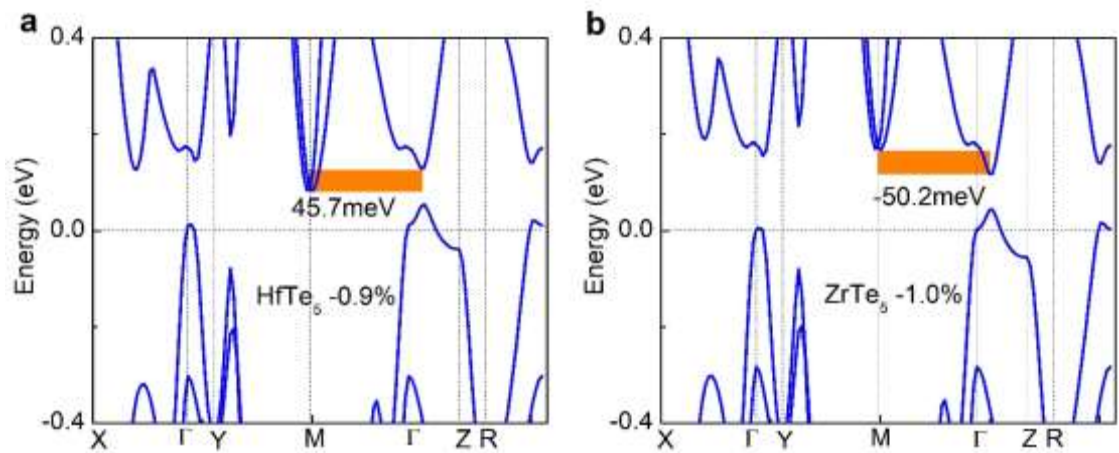

**Figure 3 | The comparison of band structure between HfTe₅ and ZrTe₅. a,** The calculated HfTe₅ band structure based on the experimental lattice constant at 10K by WIEN2k. The -0.9% means that the volume of bulk HfTe₅ at 10K is 0.9% smaller than 293K. The energy difference between Γ point and M point is 45.7meV which is marked by the orange range. **b,** Corresponding calculated ZrTe₅ band structure based on experimental lattice constant at 10K by WIEN2k whose volume is 1.0% smaller than 293K. The energy difference between Γ point and M point is -50.2meV.

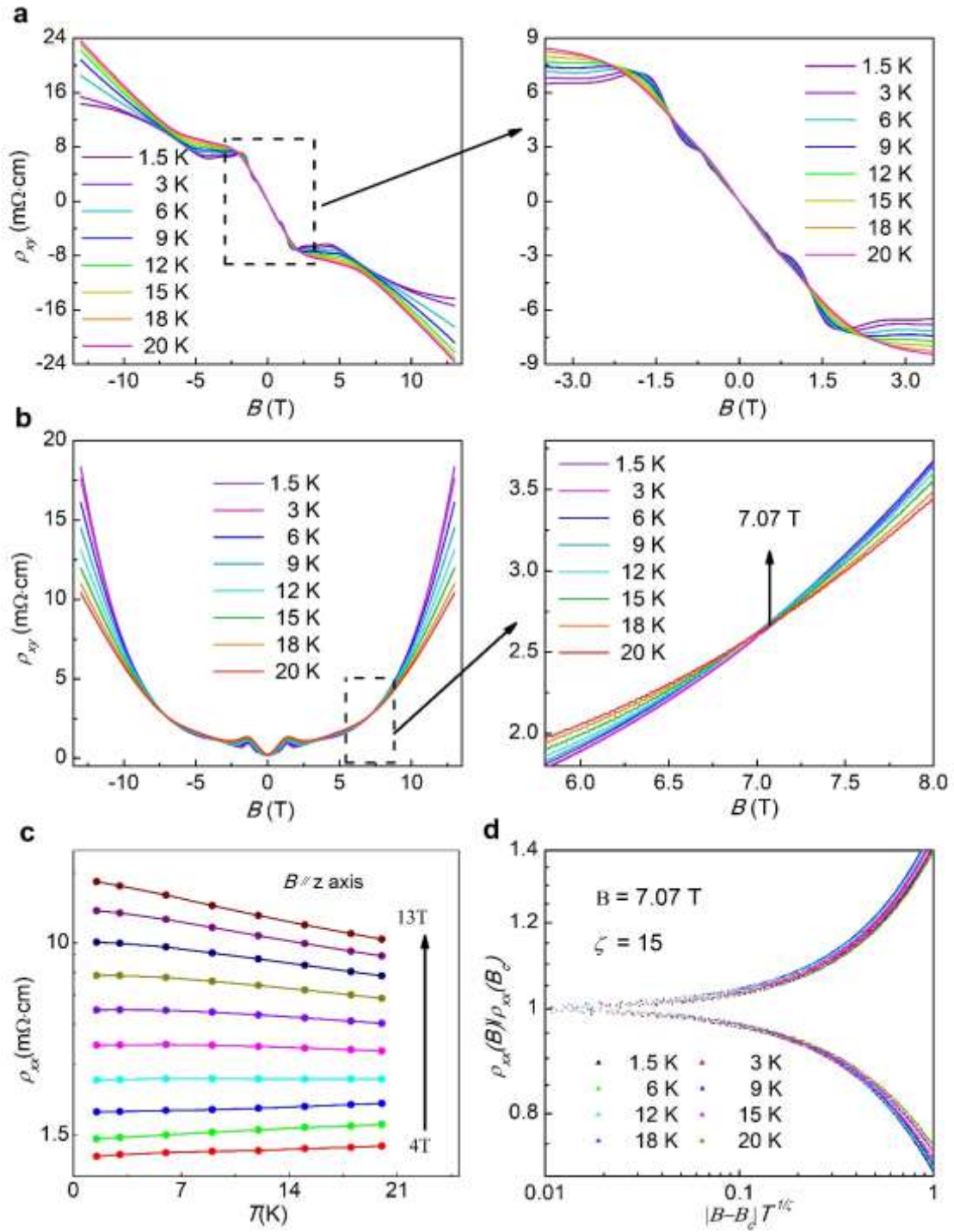

**Figure 4 | High magnetic feld-dependent of longitudinal and Hall resistivity. a,** The magnetic field-dependence of Hall resistivity $\rho_{xy}$ at different temperature T variable from 1.5 K to 20 K. **b**, The magnetic field-dependence of longitudinal resistivity $\rho_{xx}$ at different temperature $T$ variable from 1.5 K to 20 K. The right is the zoom in image of the transition cross point at 7.07 T marked by box in left figure. **c**, The temperature-dependence of longitudinal resistivity $\rho_{xx}$ at different magnetic field $B$ variable from 4

T to 13 T. We can find the tendency of the curve is different between low field to high field. **d**, The Normalized resistance $\rho_{xx}/\rho_{xx}(B_c)$ as a function of the scaling variable $|B-B_c|T^{-1/\zeta}$ at indicated temperatures. The best fitting parameter is at $\zeta=15$.